\documentclass[a4paper, 12pt, oneside]{article}

\usepackage{aapreprint}

\usepackage{graphicx,wrapfig,float,slashed,subcaption,bbold,bm}
\usepackage{amsmath,amssymb,epsfig,graphicx,xcolor}
\usepackage{epstopdf}
\usepackage{ragged2e}
\usepackage{mciteplus}

\newcommand{\nc}{\newcommand}
\nc{\non}{\nonumber}
\nc{\hc}{\hbox {H.c.}}
\nc{\noi}{\noindent}
\nc{\barx}{\bar{x}}
\nc{\pbarn}{\;\hbox {pb}}
\nc{\fbarn}{\;\hbox {fb}}

\nc{\hsp}{\hspace{0.5cm}}
\nc{\lsp}{\hspace{1cm}}
\nc{\Lsp}{\hspace{2cm}}
\nc{\LLsp}{\lsp\lsp}
\nc{\lra}{\longrightarrow}
\nc{\p}{\prime}
\nc{\sgn}{\text{sgn}}
\nc{\ph}{\varphi}
\nc{\op}{{\cal O}}

\nc{\beq}{\begin{equation}}  \nc{\eeq}{\end{equation}}
\nc{\bea}{\begin{eqnarray}}  \nc{\eea}{\end{eqnarray}}
\nc{\baa}{\begin{array}}     \nc{\eaa}{\end{array}}
\nc{\bit}{\begin{itemize}}   \nc{\eit}{\end{itemize}}
\nc{\ben}{\begin{enumerate}} \nc{\een}{\end{enumerate}}
\nc{\bce}{\begin{center}}    \nc{\ece}{\end{center}}
\nc{\bpm}{\begin{pmatrix}}   \nc{\epm}{\end{pmatrix}}
\nc{\bvt}{\begin{verbatim}}  \nc{\evt}{\end{verbatim}}

\def\lsim{\mathrel{\raise.3ex\hbox{$<$\kern-.75em\lower1ex\hbox{$\sim$}}}}
\def\gsim{\mathrel{\raise.3ex\hbox{$>$\kern-.75em\lower1ex\hbox{$\sim$}}}}

\def\udots{\mathinner{\mkern1mu\raise1pt\vbox{\kern7pt\hbox{.}}\mkern2mu\raise4pt\hbox{.}\mkern2mu\raise7pt\hbox{.}\mkern1mu}}

\def\kev{\;\hbox{keV}}

\allowdisplaybreaks

\newcommand\fverb{\setbox\fverbbox=\hbox\bgroup\verb}
\newcommand\fverbdo{\egroup\medskip\noindent%
			\fbox{\unhbox\fverbbox}\ }
\newcommand\fverbit{\egroup\item[\fbox{\unhbox\fverbbox}]}
\newbox\fverbbox

\preprint{\begin{flushright}
UTTG 14-2019
\end{flushright}}

\title{Suppressed flavor violation in Lepton Flavored Dark Matter from an extra dimension} 
\author[a]{Niral Desai,}
\author[a]{Can Kilic,}
\author[a]{Yuan-Pao Yang,}
\author[a]{Taewook Youn}
\affiliation[a]{Theory Group, Department of Physics\\ University of Texas at Austin,
Austin, TX 78712, USA}
\emailAdd{npd393@utexas.edu}
\emailAdd{kilic@physics.utexas.edu}
\emailAdd{yjp1986@utexas.edu}
\emailAdd{taewook.youn@utexas.edu}

\abstract{Phenomenological studies of Flavored Dark Matter (FDM) models often have to assume a near-diagonal flavor structure in the coupling matrix in order to remain consistent with bounds from flavor violating processes. In this paper we show that for Lepton FDM, such a structure can naturally arise from an extra dimensional setup. The extra dimension is taken to be flat, with the dark matter and mediator fields confined to a brane on one end of the extra dimension, and the Higgs field to a brane on the other end. The Standard Model fermion and gauge fields are the zero modes of corresponding bulk fields with appropriate boundary conditions. Global flavor symmetries exist in the bulk and on the FDM brane, while they are broken on the Higgs brane. Flavor violating processes arise due to the misalignment of bases for which the interactions on the two branes are diagonalized, and their size can be controlled by a choice of the lepton profiles along the extra dimension. By studying the parameter space for the model, we show that when relic abundance and indirect detection constraints are satisfied, the rates for flavor violating processes such as $\mu\to e\gamma$ remain far below the experimental limits.}

\begin{document}

\maketitle
\flushbottom

\section{Introduction}
\label{sec:intro}

While the existence of dark matter (DM) is strongly supported by astronomical observations, its microscopic nature remains a mystery. In the absence of experimental input from particle physics experiments such as the Large Hadron Collider (LHC), direct, or indirect DM detection experiments, models of DM are designed to be simple, and to be compatible with extensions of the Standard Model that are motivated by other considerations. For instance, in models that address the naturalness problem of the scalar sector in the Standard Model (SM) by introducing partner particles that are odd under a $Z_{2}$ symmetry, the DM can be the lightest partner particle, which often leads to its observed relic abundance through thermal production in the early universe. Alternatively, models of asymmetric DM \cite{Nussinov:1985xr, Gelmini:1986zz,Barr:1990ca,Barr:1991qn,Kaplan:1991ah,Kaplan:2009ag,Petraki:2013wwa,Zurek:2013wia} allow for a simple connection between DM and the matter/antimatter asymmetry in the SM sector. Axion DM \cite{PhysRevD.16.1791,PhysRevLett.38.1440,PhysRevLett.40.223,PhysRevLett.40.279} is motivated by its connection to the strong CP problem. 

Recently, models of Flavored Dark Matter (FDM) \cite{MarchRussell:2009aq,Cheung:2010zf,Kile:2011mn,Batell:2011tc,Agrawal:2011ze,Kumar:2013hfa,Lopez-Honorez:2013wla,Kile:2013ola,Batell:2013zwa,Agrawal:2014una,Agrawal:2014aoa,Hamze:2014wca,Lee:2014rba,Kile:2014jea,Kilic:2015vka,Calibbi:2015sfa,Agrawal:2015tfa,Bishara:2015mha,Bhattacharya:2015xha,Baek:2015fma,Chen:2015jkt,Agrawal:2015kje,Yu:2016lof,Agrawal:2016uwf,Galon:2016bka,Blanke:2017tnb,Blanke:2017fum,Renner:2018fhh,Dessert:2018khu} have been introduced to consider a different type of connection, between DM and the flavor structure of the SM. In FDM models, the DM is taken to transform non-trivially under lepton, quark, or extended flavor symmetries, and it couples to SM fermions at the renormalizable level via a mediator. This coupling is taken to be of the form
\beq
{\mathcal L}\supset \lambda_{ij} \bar{\chi}_{i} \psi_{j}\phi + {\rm h.c.},
\label{eq:FDMgeneral}
\eeq
where the $\chi_{i}$ represent the DM ``flavors'', the $\psi_{j}$ are generations of a SM fermion (such as the right-handed leptons) and $\phi$ is the mediator. Both particle physics as well as astrophysical signatures of FDM have become active areas of research.

Because of the non-trivial flavor structure of the interaction of equation~\ref{eq:FDMgeneral}, one of the main phenomenological challenges for FDM models is to keep beyond the Standard Model flavor changing processes under control. Indeed, when no specific structure is assumed for the entries in the $\lambda_{ij}$ matrix, the off-diagonal elements can give rise to flavor changing neutral currents (FCNCs) with rates that are excluded experimentally~\cite{TheMEG:2016wtm,Tanabashi:2018oca}. Most phenomenological studies of FDM models simply assume that the entries in the $\lambda_{ij}$ matrix have a specified form, such as Minimal Flavor Violation (MFV)~\cite{DAmbrosio:2002vsn}, in order to minimize flavor violating processes, but it is not clear that there is a UV completion of the FDM model where the MFV structure arises naturally. 

In this paper we will adopt a benchmark of lepton-FDM, where the SM fields participating in the FDM interaction of equation~\ref{eq:FDMgeneral} are the right-handed ($SU(2)$ singlet) leptons, and we will show that in a (flat) five-dimensional (5D) UV completion\footnote{For a comprehensive review of extra dimensional models, see for instance ref.~\cite{Ponton:2012bi} and references therein.} of this model, the rates of flavor violating processes can be naturally small. In fact, as we will show, in the region of parameter space where relic abundance and indirect detection constraints are satisfied, the branching fraction for $\mu\rightarrow e\gamma$, which is the leading flavor violating process, is orders of magnitude below the experimental bounds. 

We take the DM ($\chi_{i}$) and mediator ($\phi$) fields to be confined to a brane on one end of the extra dimension (the ``FDM brane''), and the Higgs field to be confined to a brane on the other end (the ``Higgs brane''), while the SM fermion and gauge fields are the zero modes of corresponding 5D bulk fields. In the bulk and on the FDM brane, there exist global $SU(3)$ flavor symmetries for each SM fermion species $\{q_{L}, u_{R}, d_{R}, \ell_{L}, e_{R}\}$, but these symmetries are broken on the Higgs brane. Flavor violation can only arise due to the mismatch between the basis in which the Yukawa couplings and the boundary-localized kinetic terms (BLKTs) \cite{Georgi:2000ks,Dvali:2001gm,Carena:2002me,delAguila:2003bh,delAguila:2003kd,delAguila:2003gv,delAguila:2006atw} on the Higgs brane are diagonal, and the basis in which the interaction of equation~\ref{eq:FDMgeneral} on the FDM brane is diagonal. Naively, one may think that no such mismatch can arise, since the FDM interaction starts out proportional to $\delta_{ij}$, and must therefore remain so after any unitary basis transformation. The Higgs brane BLKTs however cause shifts in the normalization of the lepton kinetic terms in a non-flavor universal way, and therefore the basis transformation necessary to bring the fields back into canonically normalized form involves rescalings, which are not unitary. By the time this is done and all interactions on the Higgs brane are brought to diagonal form, the FDM interaction is no longer diagonal. However, the size of the off-diagonal entries can be controlled by adjusting the profiles of the leptons along the extra dimension. In particular, by an appropriate choice of bulk masses, the fermion profiles can be made to peak on either brane, and be exponentially suppressed on the other. In the limit where the lepton profiles are sharply peaked on the FDM brane, the effect of all Higgs brane couplings vanish, and there is no flavor violation. Of course, in that limit the lepton zero modes, which only obtain masses from the Yukawa interactions on the Higgs brane, also become massless. Thus there is a tension between reproducing the correct lepton masses and suppressing lepton flavor violating processes. In the rest of this paper, we will quantitatively study this setup, and show that there are regions in the parameter space where the model can be made consistent with all experimental constraints.

The layout of the paper is as follows: In section~\ref{sec:xdim}, we will introduce the details of the 5D model. Then in section~\ref{sec:constraints}, we will study the impact of constraints from relic abundance, direct and indirect DM detection experiments, flavor violating processes and collider searches on the parameter space of the model. We will conclude in section~\ref{sec:conclusions}.

\section{Details of the model}
\label{sec:xdim}

\noindent
{\bf Generalities:} As described in the introduction, we will adopt a benchmark model of lepton-FDM. Since we wish to consider a 5D UV completion, it is convenient to make use of 4 component Dirac spinor notation. We introduce three flavors of DM
\beq
\Psi_{\chi,i}=\left(\begin{array}{c} \chi_{L,i}\\ \chi_{R,i} \end{array}\right),
\eeq
and a scalar mediator field $\phi$ with hypercharge $+1$, such that the 4D effective Lagrangian contains an interaction between $\chi_{L,i}$ and the right handed leptons $e_{R,j}$
\beq
{\mathcal L}\supset \lambda_{ij} \bar{\chi}_{L,i} e_{R,j}\phi + {\rm h.c.}.
\label{eq:FDM}
\eeq

This effective interaction arises from an orbifolded flat extra dimension of length $L$, with the FDM brane at $y=0$ and the Higgs brane at $y=L$. As we will see in section~\ref{sec:constraints}, constraints on the resonant production of the Kaluza-Klein (KK) modes of the SM gauge bosons suggest that the KK scale must be $\pi / L \gsim 10~$TeV, but we remark that the KK scale can in principle be much higher ($L^{-1}\lsim M_{\rm Planck,5D}$), which significantly simplifies the cosmological history. We will make no further assumptions about the KK scale.

\noindent
{\bf Field Content:} The SM gauge fields and fermions will all be taken to be the zero modes of corresponding 5D fields in the bulk. The boundary conditions for these fields are chosen such that the chiral matter content of the SM arises in the zero modes \cite{Ponton:2012bi}. In particular, we introduce
\beq
\Psi_{\ell,i}=\left(\begin{array}{c} \ell_{L,i}\\ \ell_{R,i} \end{array}\right)\qquad {\rm and}\quad \Psi_{e,i}=\left(\begin{array}{c} e_{L,i}\\ e_{R,i} \end{array}\right).
\eeq
Where the SM left-handed ($SU(2)$ doublet) and right-handed ($SU(2)$ singlet) leptons are the zero modes of $\ell_{L,i}$ and $e_{R,j}$, while the zero modes of $\ell_{R,i}$ and $e_{L,j}$ are projected out by the boundary conditions. Additionally, the left-handed quark doublet and two quark singlets 
\beq 
\Psi_{q,i} = \left(\begin{array}{c} q_{L,i}\\ q_{R,i} \end{array}\right) \qquad \Psi_{u,i} = \left(\begin{array}{c} u_{L,i}\\ u_{R,i} \end{array}\right) \qquad \Psi_{d,i} = \left(\begin{array}{c} d_{L,i}\\ d_{R,i} \end{array}\right)
\eeq 
exist in the bulk, and the zero modes of $q_{R,i}, u_{L,i},$ and $d_{L,i}$ are projected out similarly to the leptons. The boundary conditions for the SM gauge bosons are chosen such that the $A^{5}$ is projected out for all of them.

The Higgs doublet field $H$ is taken to be confined to the Higgs brane, where the SM Yukawa couplings arise, and the FDM fields $\Psi_{\chi,i}$ and $\phi$ are taken to be confined to the FDM brane, where the FDM interaction of equation~\ref{eq:FDM} arises.

\noindent
{\bf Flavor Structure:} In our model, the bulk and FDM brane respect an exact flavor symmetry ${\mathcal G}_{lepton}=SU(3)_{\ell}\times SU(3)_{e}$ within the lepton sector, under which the $\Psi_{\ell,i}$ transform as $({\bf 3},{\bf 1})$, while the $\Psi_{e,i}$ and $\Psi_{\chi,i}$ transform as $({\bf 1},{\bf 3})$. This symmetry is broken on the Higgs brane. Consequently, the lepton Yukawa couplings are not a-priori assumed to have a special flavor structure. Of course, in the absence of any other source of symmetry breaking, the Yukawa terms can be brought into diagonal form via a change of basis, and a $U(1)^{3}$ symmetry will survive, forbidding any lepton flavor violating processes. However, due to the absence of a flavor symmetry on the Higgs brane, we also need to include BLKTs for the leptons that are off-diagonal, and as we will show below in detail, together with the FDM interaction these generically break the flavor symmetry down to just the overall lepton number ($U(1)_{L}$) such that lepton flavor violating processes are no longer forbidden. As we will show in section~\ref{sec:constraints} these processes can be well below experimental bounds with a natural choice of parameters in our setup.

The quark sector has a similar flavor symmetry ${\mathcal G}_{quark} = SU(3)_q \times SU(3)_u \times SU(3)_d$ in the bulk and on the FDM brane. ${\mathcal G}_{quark}$ is also broken on the Higgs brane by Yukawa couplings and BLKTs down to overall baryon number ($U(1)_{B}$). Unlike the leptons however, there are no additional interactions for the quarks on the FDM brane, and due to gauge symmetry, the BLKTs are diagonal in the same basis as the kinetic terms. As a result, the only source of quark flavor violation in addition to those already present in the SM arises at loop level due to KK quarks in loops. Since the KK scale can be arbitrarily large, there are no further constraints from flavor violation in the quark sector.

\noindent
{\bf KK mode decomposition:} Bulk fermions have mass terms $M_\Psi$, which determine the 5D profiles of the zero modes. In particular, the profiles of the fermion zero modes are proportional to $e^{-M_\Psi x^{5}}$. We will choose the mass parameters such that the right-handed lepton profiles peak towards the FDM brane and they are suppressed at the Higgs brane. This can explain the smallness of the 4D effective $\tau$ Yukawa coupling for $\mathcal O(1)$ values of $M_\Psi L$, as we will show in Section 3. Due to the unbroken flavor symmetries in the bulk, for each of the SM fermions $\{q_{L},u_{R},d_{R},\ell_{L},e_{R}\}$, the three generations have identical profiles, thus the small {\it ratios} of Yukawa couplings $y_{e} / y_{\tau}$ and $y_{\mu} / y_{\tau}$ will not be addressed in our model. Explicitly, the KK mode decomposition for a 5D fermion field with a bulk mass $M_{\Psi}$ can be written as:
\beq
\Psi(x^{\mu},x^{5})=\frac{C_{\Psi}}{\sqrt{L}}\ e^{-M_{\Psi} x^{5}}\ \Psi^{0}(x^{\mu})+\sum_{n=1}^{\infty} f_{\Psi,n}(x^{5}) \Psi^{n}(x^{\mu}),
\label{eq:KK}
\eeq
where $\Psi^{0}$ is the zero mode, the coefficient $C_{\psi}$ is chosen such that the kinetic term for $\Psi^{0}$ is properly normalized, and the $\Psi^{n}$ are the KK modes, with profiles $f_{\Psi,n}(x^{5})$ in the extra dimension. As we will see below, the smallness of lepton flavor violating processes is a consequence of the lepton zero mode profiles being small on the Higgs brane.

\noindent
{\bf Interactions:} The bulk Lagrangian contains only the kinetic terms for the gauge fields as well as the kinetic (with minimal gauge coupling) and mass terms for the fermions. The Lagrangian of the lepton sector on the FDM brane includes, in addition to the $\Psi_{\chi}$ and $\phi$ kinetic and mass terms (the $\Psi_{\chi}$ are degenerate at this level due to the flavor symmetry $\mathcal{G}_{lepton}$), the following terms
\beq
{\mathcal L}_{y=0} \supset  \left(\lambda_{0} \delta_{ij} \overline{\Psi}_{\chi_i} \Psi_{e_{j}}  \phi + {\rm h.c.}\right) + \alpha^{\ell}_{0} \delta_{ij} \overline{\Psi}_{\ell_{i}} i \partial_{\mu} \gamma^{\mu} \Psi_{\ell_{j}} + \alpha^{e}_{0} \delta_{ij} \overline{\Psi}_{e_{i}} i \partial_{\mu} \gamma^{\mu} \Psi_{e_{j}}.
\label{eq:FDM5D}
\eeq
Keeping only the zero modes for the leptons, this becomes
\begin{eqnarray}
{\mathcal L}_{y=0}& \supset & \left(\lambda_{0}\frac{C_{e}}{\sqrt{L}} \delta_{ij} \bar{\chi}_{L,i} e_{R,j}  \phi + {\rm h.c.}\right) \nonumber\\
 & & + \alpha_{0}^{\ell}\frac{C_{\ell}^{2}}{L} \delta_{ij} \bar{\ell}_{L,i} i \partial_{\mu} \bar{\sigma}^{\mu} \ell_{L,j} + \alpha_{0}^{e} \frac{C_{e}^{2}}{L} \delta_{ij} \bar{e}_{R,i} i \partial_{\mu} \bar{\sigma}^{\mu} e_{R,j}.
\label{eq:FDMzeromodes}
\end{eqnarray}
Thus the effective size of the coupling in the FDM interaction is
\beq
\lambda\equiv \frac{\lambda_{0}}{\sqrt{L}}\ C_{e}.
\label{eq:lambda}
\eeq
Thus assuming the dimensionless quantity $\lambda_{0} / \sqrt{L}$ appearing in the 5D theory to be $\mathcal O(1)$, the 4D effective FDM coupling is not particularly suppressed. Note that for completeness we have included BLKTs on the FDM brane. However due to the exact flavor symmetry ${\mathcal G}$ there, those are characterized only by the two dimensionless quantities $\alpha_{0}^{\ell,e} / L$, and the flavor structure is proportional to $\delta_{ij}$, just like the couplings of the FDM interaction. The FDM brane BLKTs do not contribute to flavor violating processes. The only effect of $\alpha_{0}^{\ell,e} / L$ is to change the normalization coefficients $C_{\ell}$ and $C_{e}$ when bringing the zero modes to canonically normalized form, but apart from that they have no more role to play in the rest of this paper.

The Lagrangian on the Higgs brane includes, in addition to the SM Higgs kinetic term and potential, the lepton Yukawa couplings, as well as BLKTs:
\beq
{\mathcal L}_{y=L} \supset  \left(Y^{L}_{0,ij} \overline{\Psi}_{\ell_{i}} \Psi_{e_{j}} H + \mathrm{h.c.} \right) + \alpha^{\ell}_{0,ij} \overline{\Psi}_{\ell_{i}} i \partial_{\mu} \gamma^{\mu} \Psi_{\ell_{j}} + \alpha^{e}_{0,ij} \overline{\Psi}_{e_{i}} i \partial_{\mu} \gamma^{\mu} \Psi_{e_{j}}.
\eeq
Again, concentrating on the zero modes, this becomes
\begin{eqnarray}
{\mathcal L}_{y=L} &\supset&  \left(Y^{L}_{0,ij} \frac{C_{\ell}C_{e}}{L} e^{-(M_{\ell}+M_{e})L} \bar{\ell}_{L,i} e_{R,j} H + \mathrm{h.c.} \right) \nonumber\\
&&+ \alpha^{\ell}_{0,ij} \frac{C_{\ell}^{2}}{L} e^{-2M_{\ell}L} \bar{\ell}_{L,i} i \partial_{\mu} \bar{\sigma}^{\mu} \ell_{L,j} + \alpha^{e}_{0,ij} \frac{C_{e}^{2}}{L} e^{-2M_{e}L} \bar{e}_{R,i} i \partial_{\mu} \bar{\sigma}^{\mu} e_{R,j}.
\end{eqnarray}
We see that the effective Yukawa coupling becomes
\beq
Y^{L}_{ij}\equiv \frac{Y^{L}_{0,ij}}{L} \ C_{\ell}C_{e} e^{-(M_{\ell}+M_{e})L}.
\label{eq:yukcpl}
\eeq
In particular, we see that the effective 4D Yukawa couplings are down by a factor of $e^{-(M_{\ell}+M_{e})L}$ from the original (dimensionless) couplings $Y^{L}_{0} / L$ appearing in the 5D theory. Note that this is in contrast with the effective 4D FDM couplings that are unsuppressed. As mentioned earlier, this can explain the smallness of the SM $\tau$ Yukawa coupling, even with $Y^{L}_{0,\tau\tau} / L \sim \mathcal O(1)$, for $(M_\ell + M_e)L \sim \mathcal O(1)$. 

Note that the BLKT coefficients $\alpha^{\ell,e}_{0,ij}$ on the Higgs brane, unlike the BLKT coefficients $\alpha^{\ell,e}_{0}$ on the FDM brane, are not proportional to the identity (or even diagonal) in flavor space. However, the coefficients in the effective 4D theory for the zero modes are suppressed:
\beq
\alpha^{\ell}_{ij}\equiv \frac{\alpha^{\ell}_{0,ij}}{L} \ C_{\ell}^{2} e^{-2M_{\ell}L},\qquad {\rm and}\qquad \alpha^{e}_{ij}\equiv \frac{\alpha^{e}_{0,ij}}{L}\ C_{e}^{2} e^{-2M_{e}L}.
\label{eq:alphas}
\eeq
As we described in the introduction, this will play a major role in the smallness of flavor violating processes, even though the $\alpha^{\ell}_{0,ij} / L$ and $\alpha^{e}_{0,ij} / L$ coefficients may be $\mathcal O(1)$ and have no special flavor structure.

\noindent
{\bf Choice of basis:} While we have now introduced the most general Lagrangian consistent with our 5D setup and the flavor symmetry $\mathcal{G}_{lepton}$, it is not straightforward in this description to calculate the rate of flavor violating processes such as $\mu\rightarrow e \gamma$, since both the kinetic terms and the Yukawa terms (and consequently the mass terms once the Higgs field is set to its vacuum expectation value) are non-diagonal in flavor space. The description of the physics is made much simpler by performing a number of field redefinitions and rotations such that both the kinetic terms and the mass terms for the fermions become diagonal. As described in the introduction, at the end of this process, all flavor non-diagonal effects can be encoded in the FDM coupling matrix.

Let us start our discussion in a basis where the Yukawa matrix $Y^{L}$ is diagonal. $C_{\ell}$ and $C_{e}$ were chosen such that the {\it flavor-diagonal} coefficients of the kinetic terms of the zero modes are one, however due to the presence of the BLKT coefficients $\alpha^{\ell ,e}_{ij}$, there are flavor off-diagonal contributions to the kinetic terms as well. Thus as a first step, we perform $SU(3)$ rotations $U^{\ell}$ and $U^{e}$ in order to diagonalize the kinetic terms. At this point, the kinetic terms are diagonal, but not canonically normalized, so we perform rescalings on the $\Psi_{\ell,i}$ and $\Psi_{e,i}$, implemented by the (diagonal) matrices $\Delta^{\ell}$ and $\Delta^{e}$. Generically, the $\Delta^{\ell,e}$ are \textit{not} proportional to $\delta_{ij}$, due to the effects of the off-diagonal BLKT entries.

Now that the kinetic terms are diagonal in flavor space and canonically normalized, we perform another set of $SU(3)$ rotations given by the matrices $V^{\ell}$ and $V^{e}$ to bring the Yukawa interactions back into a diagonal form. At the end of this procedure only the FDM couplings are non-diagonal, and they encode all flavor-violating interactions. In going to the new basis 
\begin{eqnarray}
\Psi_{\ell,i} &\rightarrow& V^{\ell}_{ij} (\Delta^{\ell})^{-1}_{jk} U^{\ell}_{kl} \Psi_{\ell,l}\qquad {\rm and}\nonumber \\
\Psi_{e,i} &\rightarrow& V^{e}_{ij} (\Delta^{e})^{-1}_{jk} U^{e}_{kl} \Psi_{e,l},
\end{eqnarray}
the original FDM coupling matrix $\lambda\delta_{ij}$ of equation~\ref{eq:FDM5D} transforms into (suppressing flavor indices)
\beq
\lambda (U^{e})^{\dag} (\Delta^{e})^{-1} (V^{e})^{\dag}.
\label{eq:FDMnondiagonal}
\eeq
In section~\ref{sec:constraints} we will use this formula in order to estimate the size of flavor-violating processes. In particular, the size of such processes depends on off-diagonal entries of this matrix (which we will generically denote by $\delta \lambda$). In order to be consistent with constraints from lepton-flavor violating processes such as $\mu\rightarrow e \gamma$~\cite{TheMEG:2016wtm}, it is sufficient if $\delta\lambda / \lambda \lesssim \mathcal O(10^{-3})$. As we are about to describe however, there are stronger constraints on this ratio from indirect detection constraints.

\begin{figure}
\begin{center}
	\includegraphics[width=0.5\textwidth]{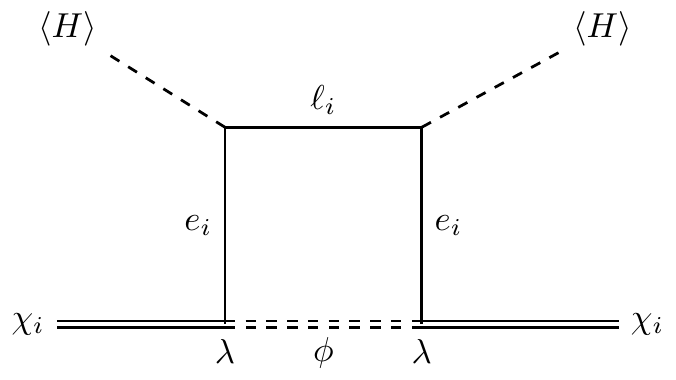}
	\caption{Flavor non-universal contribution to the $\chi$ two-point function at the one-loop level.}
	\label{fig:chimasssplitting}
\end{center}	
\end{figure}

\begin{figure}
\begin{center}
	\includegraphics[width=0.5\textwidth]{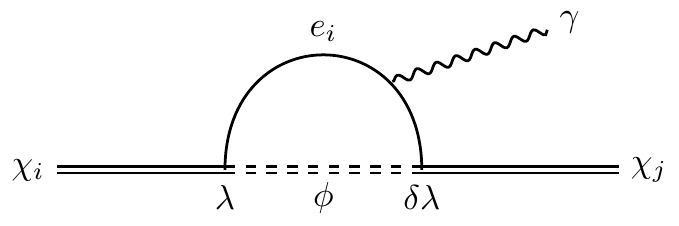}
	\caption{Leading decay mode for a heavier DM flavor to a lighter one. Note that the flavor-violating coupling $\delta \lambda$ can be on either of the vertices, depending on whether the lepton in the loop is $e_{i}$ or $e_{j}$, and the photon line can be emitted either from $\phi$ or from the lepton line.}
	\label{fig:chidecay}
\end{center}	
\end{figure}

\noindent
{\bf DM spectrum:} Note that the three $\chi$ flavors start out having degenerate masses due to the $SU(3)$ flavor symmetry on the FDM brane. However, at loop level, the breaking of the flavor symmetry is communicated to the $\chi$ fields through the diagram shown in figure~\ref{fig:chimasssplitting}. The contribution from the zero modes of the leptons in the loop is by far the dominant contribution (even when the KK scale is taken as low as 10~TeV), and since the lepton zero modes are chiral, the diagram involves $\chi_{L}$ on both sides, making this a contribution to the kinetic term (as opposed to the mass term) for $\Psi_{\chi}$. Due to the difference between the lepton Yukawa couplings, one then needs to perform flavor-dependent rescalings on the $\Psi_{\chi}$ to bring them back to canonical normalization, which induces small mass splittings. This exact mechanism leading to a mass splitting between different DM flavors was studied in ref.~\cite{Agrawal:2015tfa}, with the result that $\chi_{e}$ is the lightest flavor, and the mass splittings between flavors $i$ and $j$ are given by
\beq
\frac{\Delta m_{ij}}{m_{\chi}}=\frac{\lambda^{2}(y_{i}^{2}-y_{j}^{2})}{64\pi^{2}}\frac{v^{2}}{m_{\phi}^{2}}\left(\frac{1}{2}+\frac{m_{\chi}^{2}}{3m_{\phi}^{2}}+{\mathcal O}(\frac{m_{\chi}^{4}}{m_{\phi}^{4}})\right),
\label{eq:chisplit}
\eeq
where the $y_{i}$ are the SM lepton Yukawa couplings. For $m_{\chi}$ and $m_{\phi}$ in the TeV range, this leads to $m_{\chi,\mu}$ being larger than $m_{\chi,e}$ by $\sim \mathcal O(10)$ eV and to $m_{\chi,\tau}$ to be larger by $\sim \mathcal O(1)$ keV. The $\chi_e$ is stable as the lightest flavor. However, the heavier flavors can decay down to the lightest one through a dipole transition. This is shown in figure~\ref{fig:chidecay}. Due to the larger mass splitting and thus less phase space suppression, the $\chi_{\tau}$ lifetime is much shorter compared to the $\chi_{\mu}$ lifetime, and bounds on the production of keV-range X-rays~\cite{Ng:2019gch,Perez:2016tcq} in this decay place severe constraints on the model, $\delta\lambda / \lambda \lsim \mathcal O(10^{-6})$. Note that this is a significantly stronger constraint than the one imposed by $\mu\rightarrow e\gamma$, therefore once the X-ray indirect detection bounds are satisfied, flavor violating processes are automatically safe. We will now analyze these and other constraints on our model quantitatively.

\section{Constraints}
\label{sec:constraints}

Let us now turn our attention to the parameter space of our model, and study the impact of various types of experimental constraints on this parameter space. There are three bulk parameters of interest in the lepton sector: the KK scale $\pi / L$, and the dimensionless quantities for the bulk lepton masses $M_{\ell}L$ and $M_{e}L$. As we will see in this section, the only constraint on the KK scale arises from KK resonance searches at the LHC, while all other constraints we discuss are imposed on the dimensionless parameters of the form $M L$. On the FDM brane, we have the two masses $m_{\chi}$ and $m_{\phi}$, as well as the coupling $\lambda$. As we will describe soon, $\lambda$ will always be chosen such that the correct DM relic abundance is obtained. On the Higgs brane, we have the Yukawa couplings $Y^{L}_{0,ij}/L$ and the BLKT coefficients $\alpha^{\ell,e}_{0,ij}/L$. In the basis where the Yukawa matrix is diagonal, for a given choice of bulk masses we will choose its eigenvalues such that the correct SM lepton masses are obtained. To study the effects of the $\alpha^{\ell,e}_{0,ij}/L$ parameters, we will perform Monte Carlo studies where each element is chosen randomly, subject to positivity conditions for the matrix.

Note that since the DM and the mediator $\phi$ are confined to the FDM brane, most dark matter constraints are insensitive to the details of the extra-dimensional model. That is, they only depend on $m_\chi$, $m_\phi$, and the effective 4D FDM coupling $\lambda$, which we will fix according to the relic abundance constraint. The 5D parameters are constrained only by the bounds imposed by flavor physics, as well as the decays of the heavier DM flavors, since both processes depend on the off diagonal couplings $\delta\lambda$, the size of which is set by how suppressed the lepton profiles are on the Higgs brane.

\subsection{Dark Matter Related Constraints}

\begin{figure}
\begin{center}
	\includegraphics[width=0.45\textwidth]{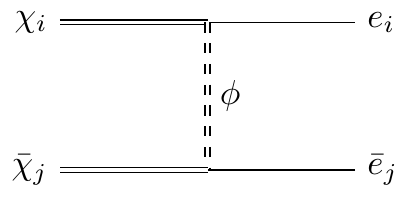}
	\caption{Dominant annihilation channel in our model.}
	\label{fig:fdmannih}
\end{center}
\end{figure}

\begin{figure}
\begin{center}
	\includegraphics[width=0.7\textwidth]{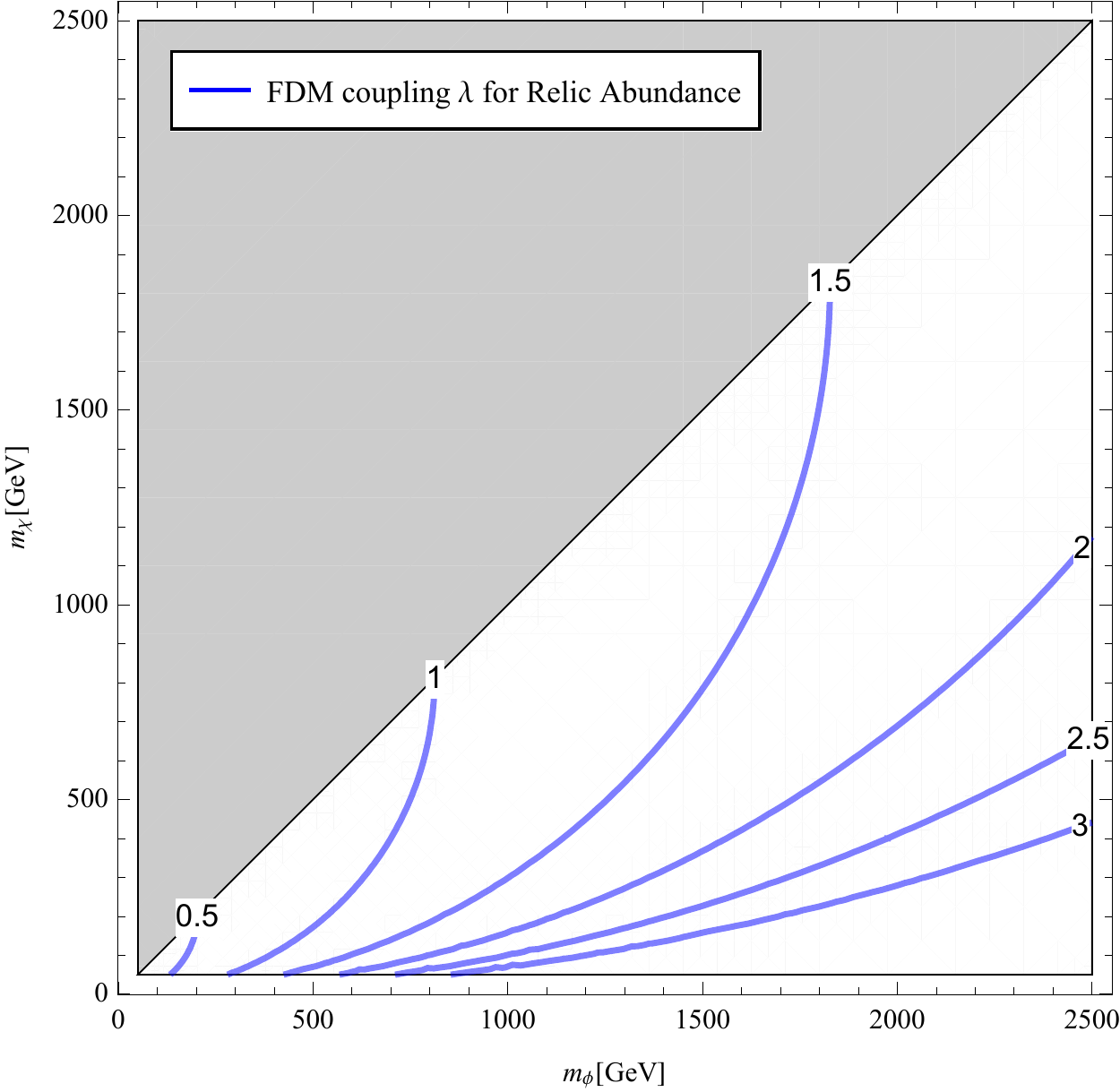}
	\caption{Contours for the value of $\lambda$ necessary to satisfy the relic abundance constraint. In the shaded region, $m_{\chi}>m_{\phi}$, and the DM is unstable.}
	\label{fig:fdmrelic}
\end{center}	
\end{figure}

{\bf Relic Abundance:} Since the three DM flavors in our model are very nearly degenerate, the relic abundance calculation closely mirrors that considered in ref.~\cite{Agrawal:2015tfa}. In particular, all three flavors freeze out at the same time, and for purposes of setting the relic abundance, they behave like a single Dirac fermion DM species. The leading annihilation process diagram is shown in figure~\ref{fig:fdmannih}. The DM abundance after freezeout has approximately equal parts $\chi_{e}$, $\chi_{\mu}$ and $\chi_{\tau}$. The annihilation cross section is given by 
\begin{equation}
\begin{aligned}
\langle \sigma v \rangle = \frac{\lambda^4 m_{\chi}^2}{32 \pi (m_{\chi}^2 + m_\phi^2)^2}.
\end{aligned}
\end{equation}
Setting $ \langle \sigma v \rangle = 2 \times (2.2 \times 10^{-26} \,\text{cm}^3\, \text{s}^{-1})$ required to obtain the correct relic abundance for a Dirac fermion, we show in figure~\ref{fig:fdmrelic} the value of $\lambda$ that is required for a range of values of $m_{\chi}$ and $m_{\phi}$. While the value is $\mathcal O(1)$ for the parameter space of interest, it is not so large that perturbative control is lost. In the rest of the paper, for any value of $m_{\chi}$ and $m_{\phi}$, we will choose $\lambda$ to equal the value that gives the correct relic abundance.

\vspace{0.2in}
\noindent
{\bf Indirect Detection:} There are two types of indirect detection signatures in our model: annihilation processes which produce leptons (as well as gamma rays from bremsstrahlung), and flavor-violating heavy $\chi_{\tau}$ decays which produce X-ray photons ($\chi_{\mu}$ decays have a much lower rate, and produce photons in the UV-range where backgrounds are much larger). Limits from the flavor-violating decays place strong bounds on the extra-dimensional parameters of this model, while the annihilation processes are insensitive to details of the extra dimension.

As mentioned in section~\ref{sec:xdim} (see figure~\ref{fig:chimasssplitting} and equation~\ref{eq:chisplit}), the mass splittings between the $\chi$'s are induced by loop processes. Specifically, one finds~\cite{Agrawal:2015tfa} that the dominant decay mode is the one shown in figure~\ref{fig:chidecay}, with one flavor-preserving coupling $\lambda$ and one flavor-violating coupling $\delta\lambda$. The width for this decay mode is given by~\cite{Agrawal:2015tfa}
\beq
\Gamma_{\Psi_{\chi,i}\rightarrow\Psi_{\chi,j}\,\gamma}=\frac{\alpha_{\rm EM}\lambda^{2}\delta\lambda^{2}}{256\pi^{4}}\frac{(\Delta m_{ij})^{3}m_{\chi}^{2}}{m_{\phi}^{4}}.
\label{eq:chidecay}
\eeq
Constraints from searches for X-rays in the keV range place bounds on the dark matter decay lifetime. These bounds are worked out in ref.~\cite{Ng:2019gch} for much lighter DM particles decaying according to the mode $\chi_{\nu}\rightarrow\gamma\nu$, with the result
\beq
\tau_{\nu} \simeq (10^{26 - 29} \mathrm{sec} ) \frac{\mathcal O(1) \kev}{m_{\chi_{\nu}}}.
\label{eq:chidecay}
\eeq
Since the decay mode in our model is $\chi_{i}\rightarrow\chi_{j}\gamma$, and the $\chi$ have ${\mathcal O}({\rm TeV})$ (as opposed to ${\mathcal O}({\rm keV})$) masses, and therefore a much lower number density, the numbers above need to be modified. In particular, the bound is on the number of photons emitted per unit time, which according to an exponential decay law during a time interval $\Delta t$ is given by $\Delta n = - (n / \tau) e^{-t/\tau} \Delta t$, where $n$ is the DM number density at time $t$. Matching this rate between the model used in refs.~\cite{Perez:2016tcq, Ng:2019gch} (with DM mass $m_{\nu}$ and lifetime $\tau_{\nu}$) and our model, and taking into account that in our model the decaying $\chi_{\tau}$ only comprise $1/3$ of the DM number density, we can set up a correspondence between the bounds on the DM lifetime in the two models:
\beq
\frac{1}{3m_{\chi_\tau} \tau_{\chi_\tau}} e^{-t_0/\tau_{\chi_\tau}} \simeq \frac{1}{m_{\nu} \tau_\nu} e^{-t_0/\tau_{\nu}},
\eeq 
$t_0 \simeq 4 \times 10^{17}$ seconds being the present age of the universe. This gives a bound on the $\chi_\tau$ lifetime of $\mathcal O(10^{17-20})$ seconds. Note that this is close to the age of the universe; in other words, if the parameters are chosen close to the bound, the $\chi_{\tau}$ particles in the universe would be just about to start decaying today in sizable numbers. For $m_\chi$ and $m_\phi$ at the TeV scale and $\lambda$ of $\mathcal O(1)$, the relevant off-diagonal coupling is constrained to be $\delta \lambda \lesssim 10^{-6}$. In section~\ref{s.otherConst} where we will perform a Monte Carlo study scanning over the BLKT coefficients, we will present distributions for the relevant off-diagonal $\lambda$ entries and we will discuss the impact on the allowed parameter space.

\begin{figure}
	\begin{center}
		\includegraphics[width=0.7\textwidth]{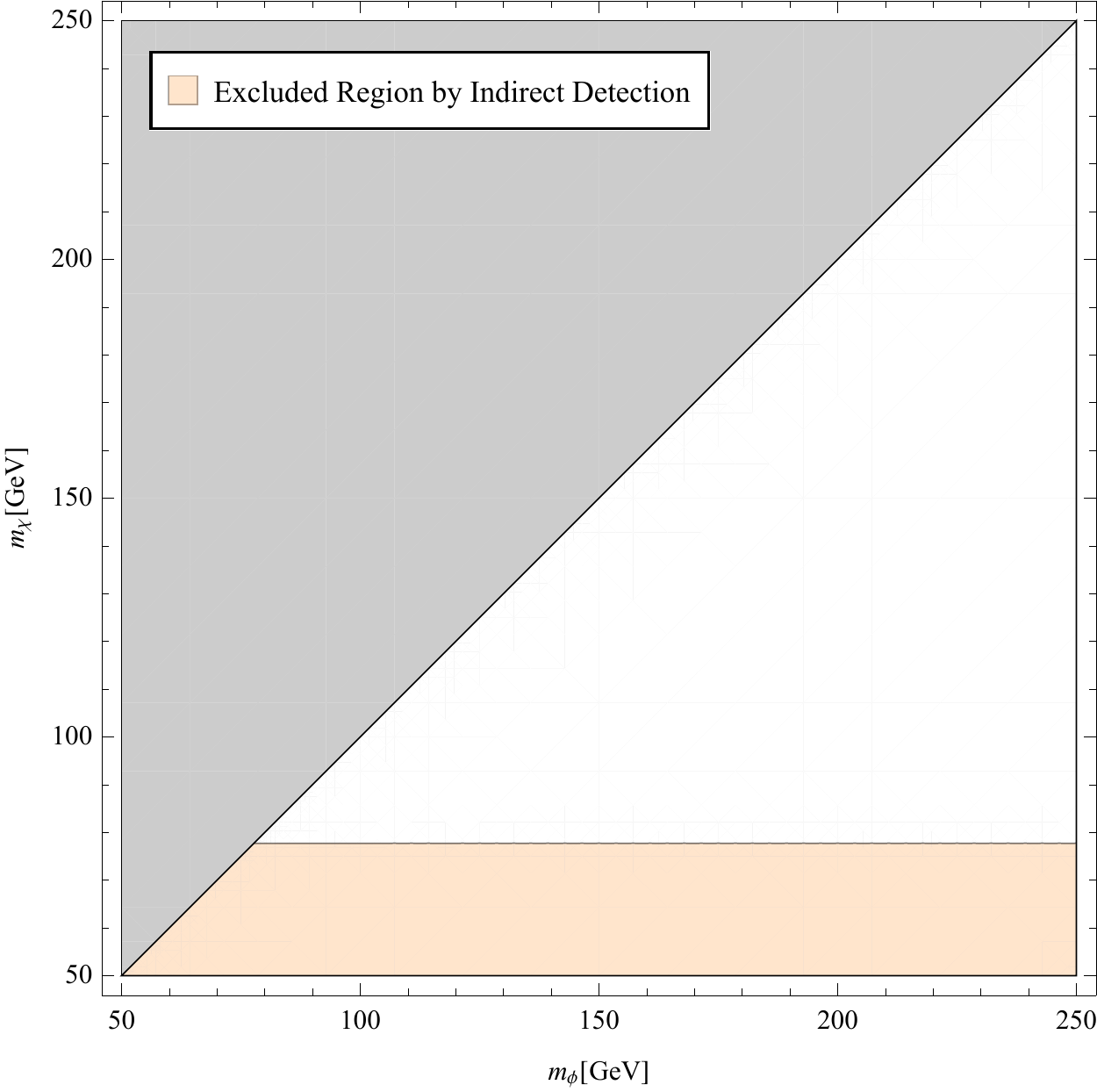}
		\caption{Indirect detection constraints on our model, with the yellow-shaded region being excluded. The most stringent bound comes from  AMS-02 experiment~\cite{Elor:2015bho}.}
		\label{fig:fdmindirect}
	\end{center}
\end{figure}

We now turn our attention to the annihilation process. The DM particles annihilate to lepton-antilepton pairs of all flavors via processes like that shown in figure~\ref{fig:fdmannih}. This leads to positron and gamma-ray signatures that experiments such as Fermi-LAT~\cite{Fermi-LAT:2016uux}, HESS~\cite{Abdallah:2016ygi}, AMS-02~\cite{Elor:2015bho} and the Planck CMB observations~\cite{Aghanim:2018eyx} are sensitive to. High energy photons are produced mainly by $\pi^{0}\rightarrow\gamma\gamma$ coming from $\tau$'s in the final state. For the mass range of interest to us, the energy of these photons is typically not high enough for HESS to place significant constraints on our model. Positrons can be produced both directly in the annihilation, as well as from the decays of $\mu^{+}$ and $\tau^{+}$ that are produced in the annihilation, although these latter sources give rise to lower positron energies, and the constraints from AMS-02 constrain primarily the directly produced positrons. The same is also true for the Planck constraints. In figure~\ref{fig:fdmindirect}, we show the effect of these indirect detection constraints (dominated by AMS-02) on our model. As we will see next, these are subdominant to direct detection constraints.
 
 \begin{figure}
	\begin{center}
		\includegraphics[width=0.65\textwidth]{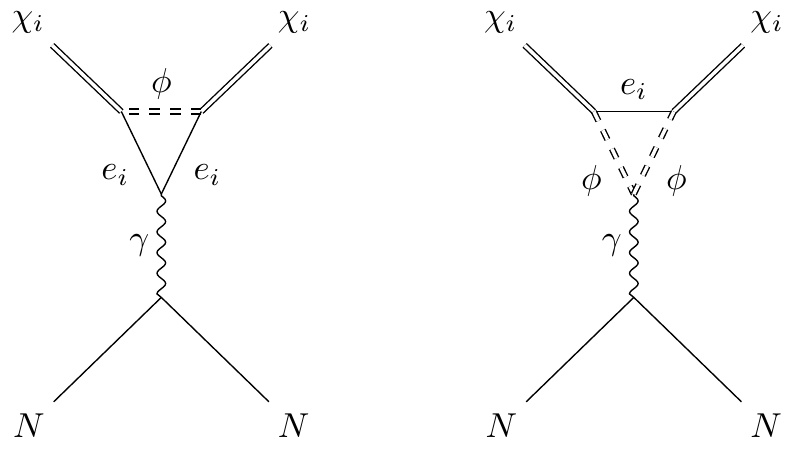}
		\caption{Leading contributions to FDM-nucleon scattering for direct detection.}
		\label{fig:fdmdirectfeyn}
	\end{center}
\end{figure}

\begin{figure}
\begin{center}
	\includegraphics[width=0.7\textwidth]{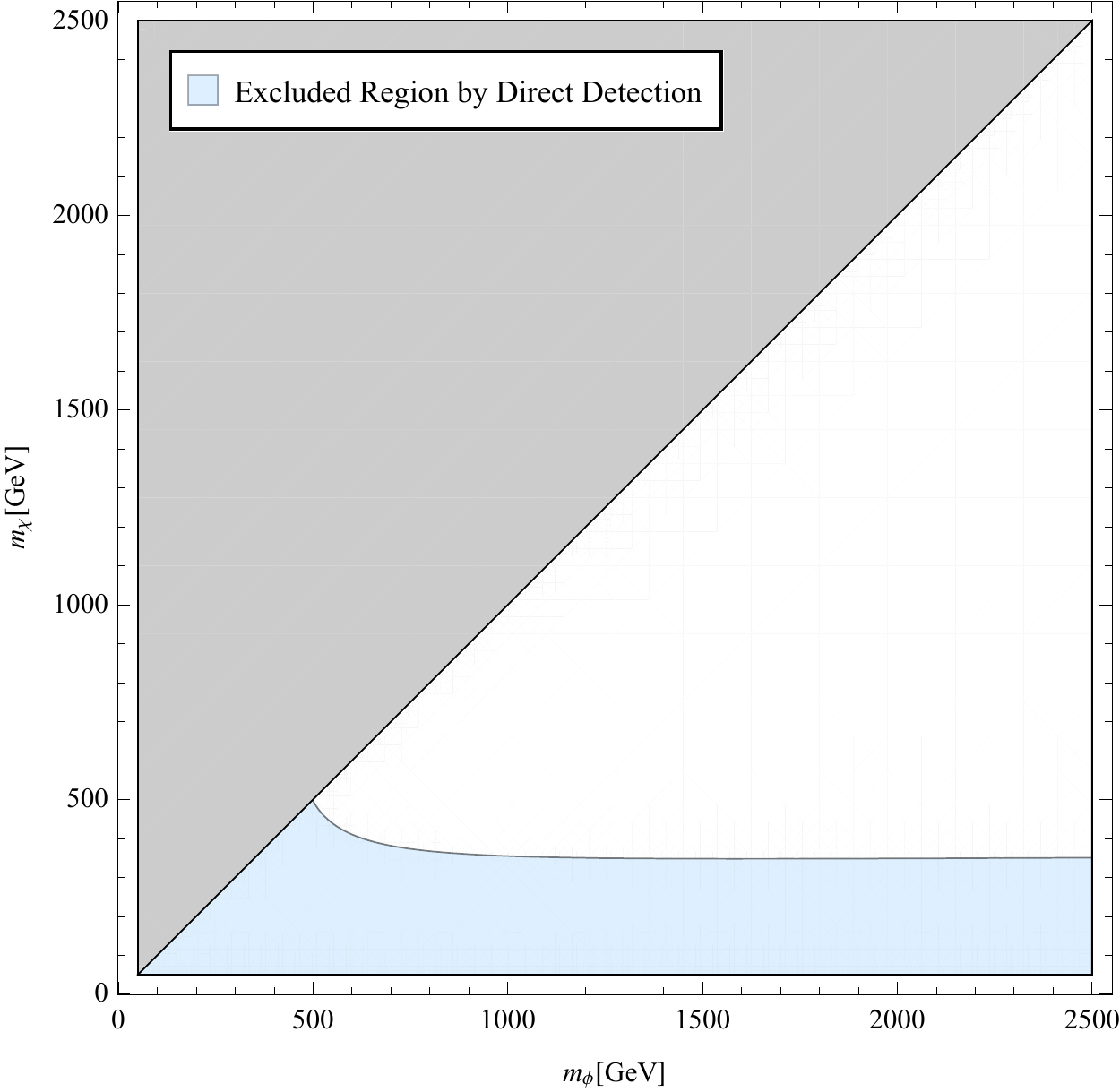}
	\caption{Direct detection constraints from the Xenon1T experiment on our model, the blue-shaded region being excluded.}\label{fig:fdmxenon}
\end{center}
\end{figure}

{\bf Direct Detection:} The scattering of one of the $\chi$ flavors from a nucleus proceeds by the loop process shown in figure~\ref{fig:fdmdirectfeyn}. The parton level cross section for flavor $i$ is given by (see \cite{Agrawal:2011ze, Hamze:2014wca})
\begin{equation}
\begin{aligned}
\sigma_{i} = \frac{\mu^2 Z^2}{\pi} \left[ \frac{\lambda^2 e^2}{64 \pi^2 m_\phi^2} \left[ 1 + \frac{2}{3}\log \left( \frac{\Lambda_{i}^{2}}{m_\phi^2} \right) \right] \right]^2,
\end{aligned}
\end{equation}
where $\mu$ is the reduced mass, and the scale $\Lambda_{i}$ cuts off the infrared divergence in the loop. For the muon and tau flavors, this scale is simply the corresponding lepton mass, $m_{\mu}$ and $m_{\tau}$ respectively. For the electron however, there is a physical scale larger than $m_{e}$ that cuts off the divergence, namely the characteristic momentum exchange in the collision. Since all three flavors of $\chi$ make up the local DM density, the effective cross section relevant for direct detection experiments is simply $(\sigma_{e}+\sigma_{\mu}+\sigma_{\tau})/3$. Using the bounds set by the Xenon1T experiment \cite{Aprile:2018dbl}, we show in figure~\ref{fig:fdmxenon} (using the value of $\lambda$ at each point that gives the correct relic abundance) the exclusion region in terms of $m_{\chi}$ and $m_{\phi}$. Values of $m_{\chi}\gsim 300$~GeV are compatible with direct detection constraints. Note that the region excluded by indirect detection constraints is already fully excluded by direct detection constraints.

\begin{figure}
\begin{center}
	\includegraphics[width=0.7\textwidth]{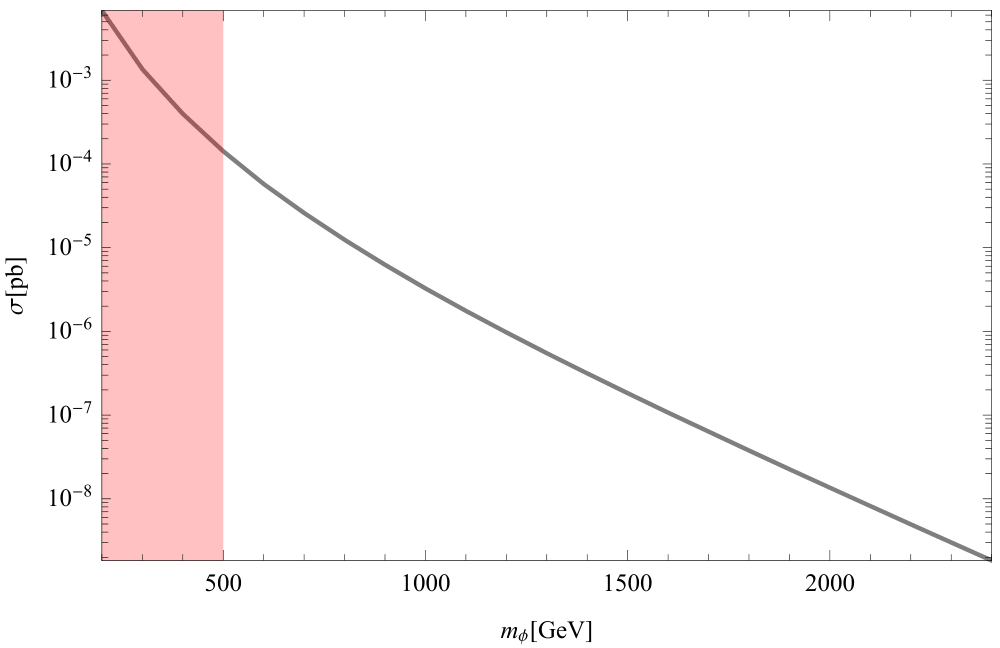}
	\caption{The $\phi$ pair-production cross section at the 13~TeV LHC as a function of $m_{\phi}$. The shaded red area denotes the values of $m_\phi$ excluded by the direct detection constraints.}
	\label{fig:fdmlhc}
\end{center}
\end{figure}

\vspace{0.2in}
\noindent
{\bf Dilepton + MET Searches at the LHC:} Since $\phi$ carries electric charge, it can be pair-produced at colliders. $\phi$ then decays to one of the three $\chi$ flavors and the associated lepton, resulting in a dilepton + MET signature where the lepton flavors on the two sides of the event are uncorrelated. The cross section for this process is very small, both since the production occurs through the electromagnetic interaction, and because $\phi$ is heavy and a scalar (which means that pair production is suppressed near threshold). We show this cross section in figure~\ref{fig:fdmlhc}, calculated with MadGraph~\cite{Alwall:2014hca}. As can be seen by comparing to figure~\ref{fig:fdmxenon}, in the region of parameter space that is not ruled out by direct detection, less than a single event is expected at existing dilepton+MET searches, and therefore even with sophisticated kinematic observables, LHC searches do not lead to any additional exclusion.

\vspace{0.2in}
\noindent

\subsection{Additional constraints}
\label{s.otherConst}

\noindent
{\bf Constraints from lepton flavor violation and DM decays:} As we have seen, constraints from X-ray searches place severe bounds on off-diagonal entries of the $\lambda$ coupling matrix. We will now work out the constraints on these couplings from lepton flavor violation, and show them to be subdominant. In the process, we will also perform a Monte Carlo study by scanning over the BLKT coefficients, and we will show the impact of the relevant constraints on the parameter space of the model.

As we have shown in section~\ref{sec:xdim}, we can work in a basis where flavor violating couplings are all encoded in the FDM coupling matrix, as in equation~\ref{eq:FDMnondiagonal}. Equations~\ref{eq:yukcpl} and~\ref{eq:alphas} imply that the effective BLKT coefficients are generally suppressed compared to the 4D Yukawa couplings. Using this fact, one can estimate the generic size of the entries of the FDM coupling matrix to the leading order in $\alpha^{\ell, e}_{ij}$ as

\begin{eqnarray}
\left[ \frac{\lambda}{\lambda_0} \right]  &\approx& 
\bordermatrix{
& (e) & (\mu) & (\tau) \cr
 (\chi_e) & 1  & \frac{ \alpha^\ell_{12} Y^L_{ee} Y^L_{\mu\mu} + \alpha^{e}_{12} (Y^L_{\mu\mu})^2}{(Y^L_{ee})^2 - (Y^L_{\mu\mu})^2}
& \frac{ \alpha^\ell_{13} Y^L_{ee} Y^L_{\tau\tau} + \alpha^{e}_{13} (Y^L_{\tau\tau})^2}{(Y^L_{ee})^2 - (Y^L_{\tau\tau})^2} \cr  (\chi_\mu) &
 \frac{ \alpha^\ell_{21} Y^L_{\mu\mu} Y^L_{ee} + \alpha^{e}_{21} (Y^L_{ee})^2}{(Y^L_{\mu\mu})^2 - (Y^L_{ee})^2}
& 1
& \frac{ \alpha^\ell_{23} Y^L_{\mu\mu} Y^L_{\tau\tau} + \alpha^{e}_{23} (Y^L_{\tau\tau})^2}{(Y^L_{\mu\mu})^2 - (Y^L_{\tau\tau})^2} \cr  (\chi_\tau) &
\frac{ \alpha^\ell_{31} Y^L_{\tau\tau} Y^L_{ee} + \alpha^{e}_{31} (Y^L_{ee})^2}{(Y^L_{\tau\tau})^2 - (Y^L_{ee})^2} 
& \frac{ \alpha^\ell_{32} Y^L_{\tau\tau} Y^L_{\mu\mu} + \alpha^{e}_{32} (Y^L_{\mu\mu})^2}{(Y^L_{\tau\tau})^2 - (Y^L_{\mu\mu})^2}
& 1} \\
&\approx& 
\left(
\begin{array}{ccc}
1  & -\mathcal O(10^{-3}) \alpha^\ell_{12} - \alpha^{e}_{12}
& -\mathcal O(10^{-4}) \alpha^\ell_{13} - \alpha^{e}_{13} \\
\mathcal O(10^{-3}) \alpha^\ell_{21} + \mathcal O(10^{-5}) \alpha^{e}_{21} 
& 1
& -\mathcal O(10^{-2}) \alpha^\ell_{23} - \alpha^{e}_{23} \\
\mathcal O(10^{-4}) \alpha^\ell_{31} + \mathcal O(10^{-8}) \alpha^{e}_{31}
& \mathcal O(10^{-2}) \alpha^\ell_{32} + \mathcal O(10^{-3}) \alpha^{e}_{32}
& 1
\end{array}
\right).
\nonumber
\label{eq:fdmapprox}
\end{eqnarray}

\begin{figure}
	\centering
	\includegraphics[width=0.5\textwidth]{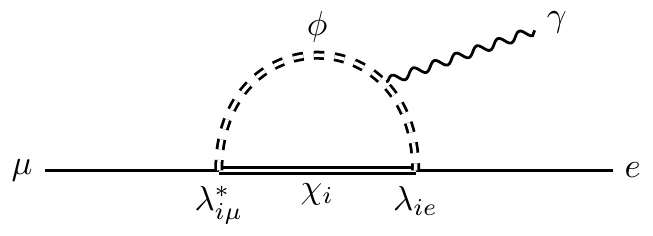}
	\caption{The leading contribution to the process $\mu\rightarrow e\gamma$ in our model. The diagrams with $\chi_\mu$ and $\chi_e$ in the loop require only one off-diagonal coupling, whereas the diagram with $\chi_\tau$ in the loop requires two off-diagonal couplings and is therefore subdominant.}
	\label{fig:mutoe}
\end{figure}

We would like to remark on a nontrivial feature of the coupling matrix. In equation~\ref{eq:FDMnondiagonal}, the matrix $V^{e}$ depends on both the $\alpha^{e}_{ij}$ as well as on $\alpha^{\ell}_{ij}$. On the other hand, the matrices $U^{e}$ and $\Delta^{e}$ depend only on $\alpha^{e}_{ij}$, but not on $\alpha^{\ell}_{ij}$, and of course they both become the identity matrix in the limit $\alpha^{e}_{ij}\to 0$. Therefore, the coupling matrix {\it squared} ($\lambda\lambda^{\dag}$ as well as $\lambda^{\dag}\lambda$) becomes identity in this limit as well, even for finite $\alpha^{\ell}_{ij}$, since $V^{e}$ is by definition a unitary matrix. It is straightforward to see from figure~\ref{fig:chidecay}, and from figure~\ref{fig:mutoe} where we show the leading contribution to the process $\mu\to e\gamma$, that for both $\chi$ decays and for lepton flavor violating processes, the FDM couplings indeed appear in the combinations $\lambda\lambda^{\dag}$ and $\lambda^{\dag}\lambda$. Therefore, for these processes, the effects of $\alpha^{\ell}_{ij}$ are always more suppressed than those of $\alpha^{e}_{ij}$, even when $\alpha^{\ell}_{ij} \gg \alpha^{e}_{ij}$, and therefore to a good approximation in the above formula, we can simply drop the $\alpha^{\ell}_{ij}$ terms.

As one can see from the expansion above, the largest elements of the coupling matrix are the $(\chi_e,\mu)$, $(\chi_e,\tau)$ and $(\chi_\mu,\tau)$ entries. Since the sizes of these entries are comparable, and since the experimental bound on the process $\mu\to e\gamma$ ($BR(\mu \to e + \gamma) < 4.2 \times 10^{-13}$ \cite{TheMEG:2016wtm}) is the strongest among lepton flavor violating processes, if this bound is satisfied in our model, then all other lepton flavor violation bounds will also be satisfied. We calculate the leading contribution to $\Gamma_{\mu\rightarrow e\gamma}$ in our model (figure~\ref{fig:mutoe}). Expanding in $m_\chi / m_\phi$, we obtain
\begin{equation}
\Gamma_{\mu\to e\gamma} = \frac{\alpha \lambda^2 (\delta \lambda)^2}{4096\pi^4}\frac{m_\mu^5}{m_\phi^4} \bigg[ \frac{1}{12}-\frac{1}{6}\frac{m_\chi^2}{m_\phi^2}+\mathcal{O}(\frac{m_\chi^4}{m_\phi^4}) \bigg]^2.
\label{eq:gammaapprox}
\end{equation} 
Here $\delta\lambda$ stands for the largest of the relevant off-diagonal couplings, which in this case is the $(\chi_e,\mu)$ element. 

In order to study the impact of both the $\mu\to e\gamma$ bound as well as the X-ray constraints from $\chi_{\tau}$ decays on the parameter space of our model, we perform a numerical study as follows: we assign random values in the interval $(-1,1)$ (sampled uniformly) to all entries of $\alpha^{\ell}_{0,ij} / L$ and $\alpha^e_{0,ij} / L$ (subject to the constraint that the kinetic terms are symmetric and positive semi-definite such that there are no ghosts in the spectrum), and we also assign random values in the interval $(0,1)$ to the FDM brane BLKT coefficients $\alpha^{e,\ell}_{0} / L$.

For any given $M_e L$ and $M_\ell L$, we can then perform the basis change procedure described in section~\ref{sec:xdim}. The values of the 5D Yukawa couplings $Y^L_{0}$ are chosen such that the correct SM lepton masses are reproduced. For a range of values for $M_e L$ and $M_\ell L$, we run 100,000 such random trials each, and we calculate the resulting  distributions for $Y^L_{0,\tau\tau}/L$ as well as all entries of the $\lambda$ matrix, and from these we calculate the $\mu\to e\gamma$ branching ratio as well as the $\chi_{\tau}$ lifetime. As a general trend, the X-ray constraints impose severe constraints on $\delta\lambda$, which as we argued above are dominated by $\alpha^{e}_{ij}$, which in turn scale as $e^{-M_{e} L}$ (see equation~\ref{eq:alphas}). Therefore the X-ray constraints favor larger values of $M_{e} L$, while they are fairly insensitive to $M_{\ell} L$.

\begin{figure}
	\centering
	\includegraphics[width=\textwidth]{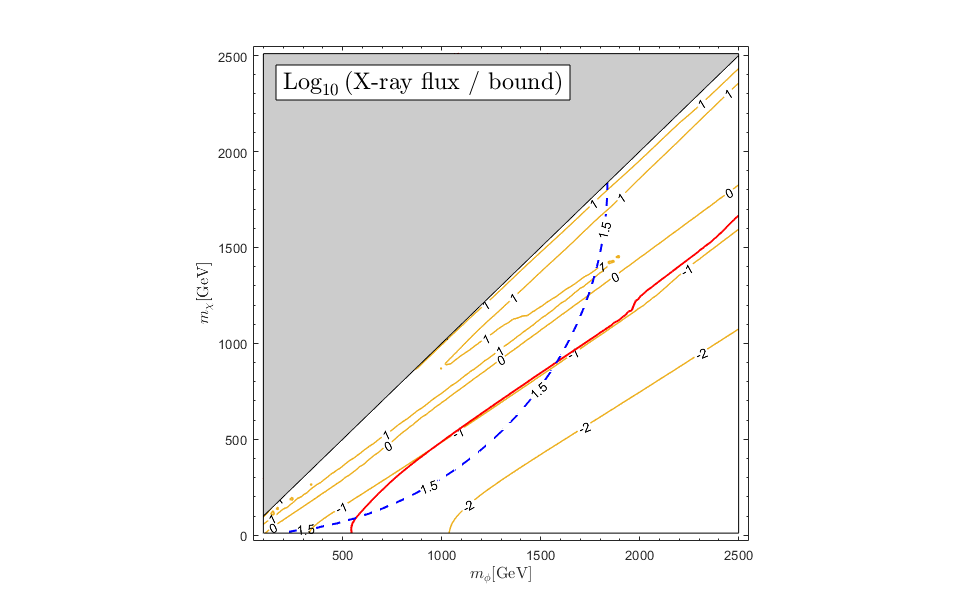}
	\caption{For the parameter point $M_e L = 8$, $M_\ell L = 1$, and as a function of $m_\chi$ and $m_\phi$, we show the contours for the ratio of the X-ray flux from $\chi_{\tau}$ decays to the limit on this flux~\cite{Perez:2016tcq, Ng:2019gch}, using the median value of $\chi_{\tau}$ lifetime obtained from 100,000 random trials. We also show in red the contour where exactly 95\% of the trials are consistent with the bound. In other words, the region to the upper left of this contour is excluded by X-ray bounds. The blue dashed curve corresponds to $\lambda=1.5$ for obtaining the correct relic abundance.}
	\label{fig:chilife}
\end{figure}

\begin{figure}
	\centering
	\includegraphics[width=1\textwidth]{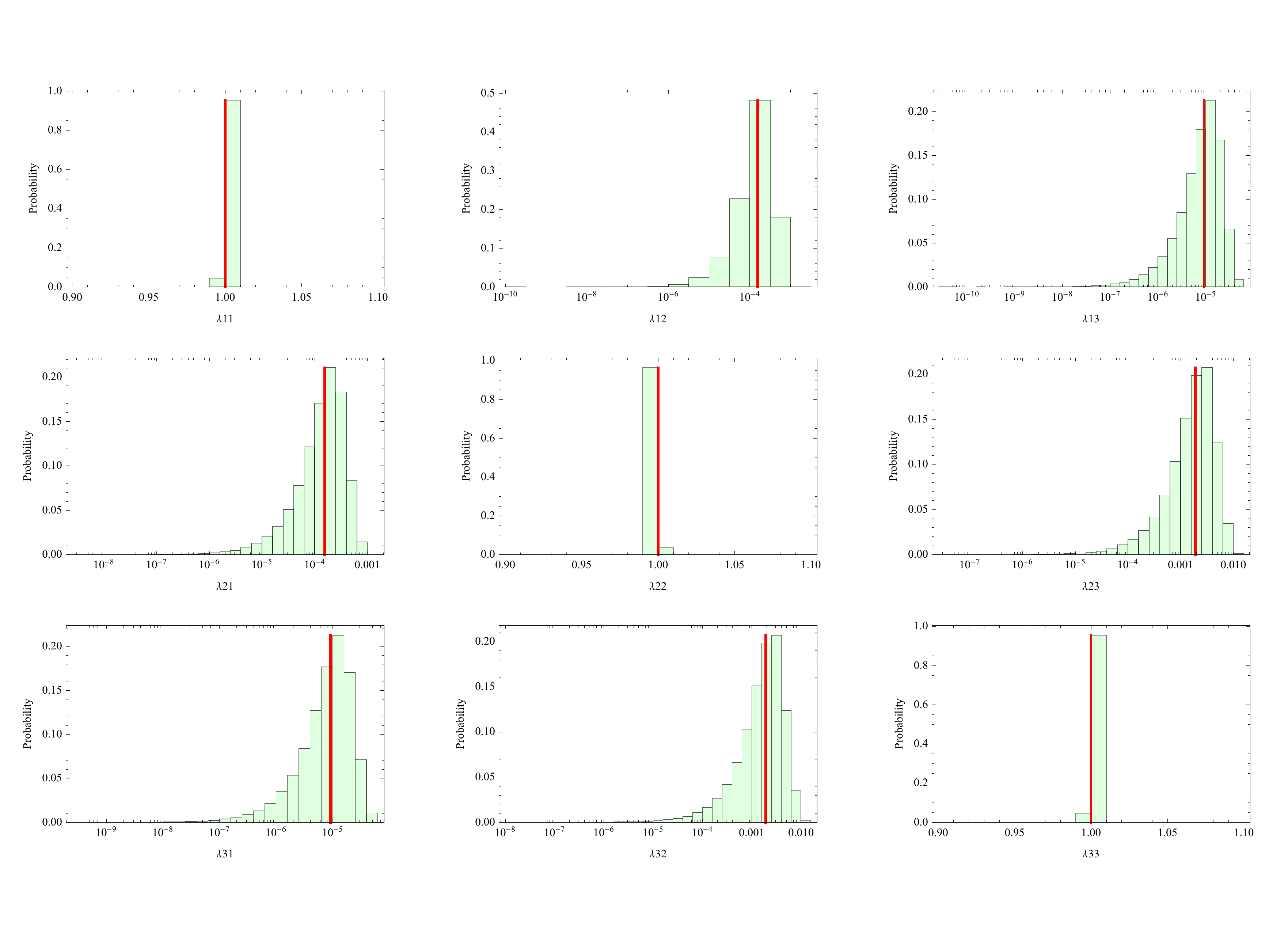}
	\caption{ For the parameter point $M_{e} L = 8$, $M_\ell L = 1$, we plot the distributions of $|\lambda_{ij}/(\frac{1}{3}\mathrm{tr}[\lambda])|$ after 100,000 random trials with $\mathcal{O}(1)$ symmetric BLKT coefficients as inputs. The analytical approximations for each entry from equation~\ref{eq:fdmapprox} are shown as red vertical lines for comparison. The first index of $\lambda_{ij}$ refers to the $\chi$-flavor, and the second index to the lepton flavor.}
	\label{fig:fdmstat}
\end{figure}

In order to gain further insight into the parameter dependence of the constraints, we show in figure~\ref{fig:chilife} how the X-ray flux from $\chi_{\tau}$ decays (where we use the median value for the $\chi_{\tau}$ lifetime from 100,000 trials) depends on $m_\chi$ and $m_\phi$, for the parameter point $M_e L = 8$, $M_\ell L = 1$. We also indicate where the exclusion contour lies, namely where exactly 95\% of trials leads to an X-ray flux consistent with the bounds~\cite{Perez:2016tcq, Ng:2019gch}. When the parameter $M_{e}$ is increased, the contours in this figure will move further up and to the left, making the excluded region smaller, while varying the parameter $M_{\ell}$ will not have a significant effect on the contours. Another way of saying this is as we increase $M_{e}$, parameter regions with smaller and smaller values of $\lambda$ become consistent with X-ray bounds. The parameter point ($M_e L = 8$, $M_\ell L = 1$) we use in this plot is chosen such that there exist points with $\lambda\lsim 1.5$ that are consistent with the bounds. For the same parameter point, we also show in figure~\ref{fig:fdmstat} the distributions of the $\lambda$ matrix entries (normalized to the diagonal entry, or more precisely $1/3$ of the trace).

\begin{figure}
	\centering
	\includegraphics[scale= 0.75]{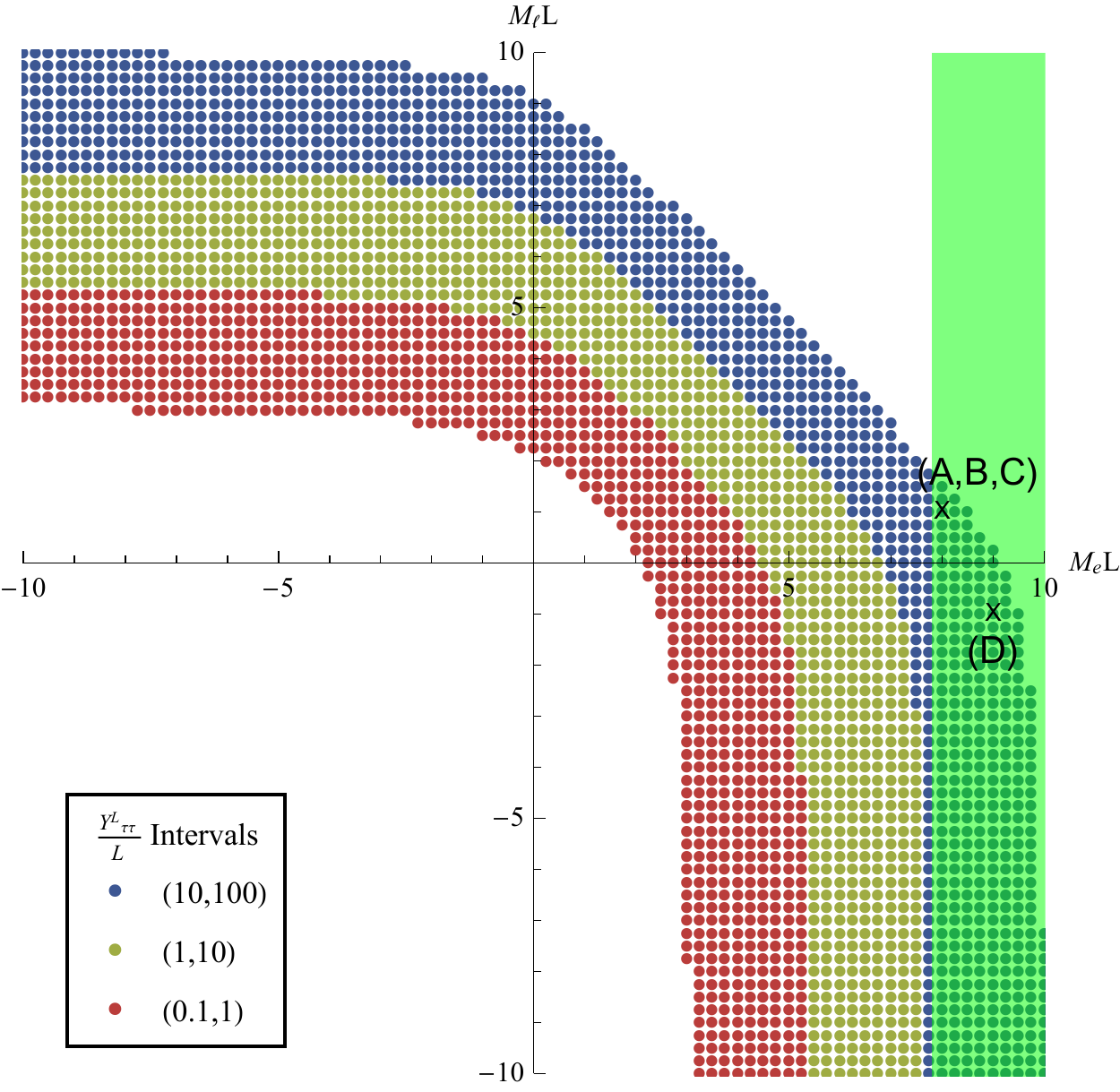}
	\caption{We show the median values of the distribution of $Y^L_{0,\tau\tau}/L$ chosen to reproduce the correct $\tau$ mass for a range of choices of $M_e L$ and $M_\ell L$, after 100,000 random trials at each point with $\mathcal{O}(1)$ BLKT coefficients as inputs. The green shaded region represents the values of $M_e L$ and $M_\ell L$ for which the model can be consistent with X-ray constraints while keeping $\lambda \le 1.5$. The parameter points A through D as indicated are used later as we study the LHC bounds on KK resonances.}
	\label{fig:Y5L}
\end{figure}

In figure~\ref{fig:Y5L}, we illustrate how the Yukawa couplings $Y^L_{0,\tau\tau}/L$ needed to reproduce the $\tau$ mass depend on the bulk mass parameters. The colors indicate in what range the median values in the distribution of $Y^L_{0,\tau\tau}/L$ lie. In this plot, we also show how the X-ray constraints impact the parameter space. In particular, in the green shaded region, there are points in the $m_\chi$-$m_\phi$ parameter space with $\lambda\lsim 1.5$ where the X-ray constraints can be satisfied. The parameter point $M_e L = 8$, $M_\ell L = 1$ used above is just inside this region. Note also that in the green shaded region, larger values of $Y^L_{0,\tau\tau}/L$ are favored.

\begin{figure}
	\centering
	\includegraphics[width=\textwidth]{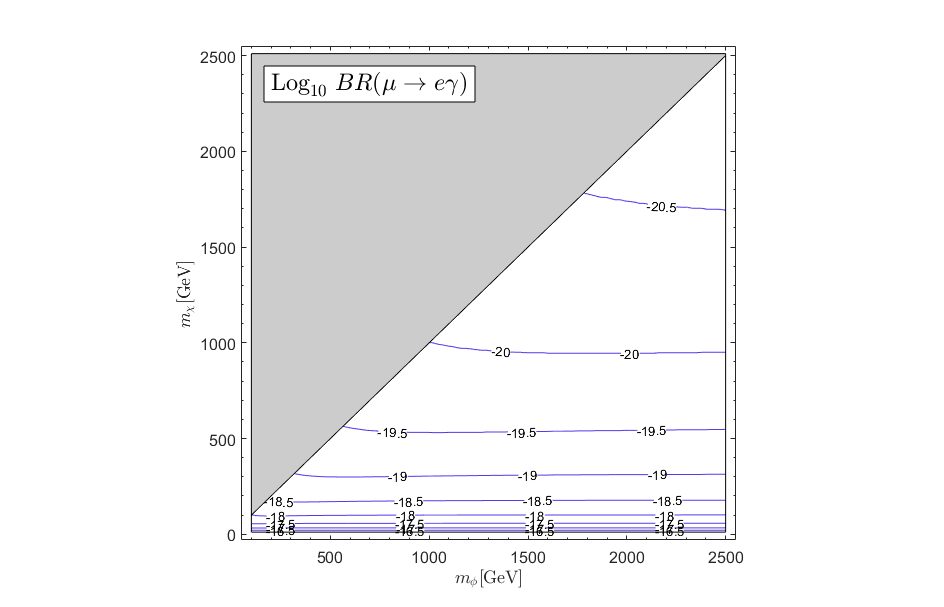}
	\caption{Contours for the median value of $\log_{10}[{\rm BR}(\mu\to e\gamma)]$ for the parameter point $M_{e} L = 8$, $M_\ell L = 1$ as a function of $m_{\chi}$ and $m_{\phi}$. For comparison, the experimental bound is BR$(\mu \to e\gamma) < 4.2 \times 10^{-13}$ \cite{TheMEG:2016wtm}.}
	\label{fig:fdmflav}
\end{figure}

Finally, also using the parameter point $M_e L = 8$, $M_\ell L = 1$ and 100,000 trials, we show in figure~\ref{fig:fdmflav} the median value of the branching ratio of $\mu\to e\gamma$ in our model as a function of $m_{\chi}$ and $m_{\phi}$. Clearly, once the X-ray constraints are satisfied, the rate of lepton flavor violating processes are far below the excluded values.

\vspace{0.2in}
\noindent
{\bf LHC Bounds on Resonant Production of KK Modes:}

In addition to the dilepton + MET final state from $\phi$ pair production and decay, another collider signature of our model is the resonant production of KK modes of the gauge bosons from $q$-$\bar{q}$ initial states (note that the triple vertex for QCD with two zero mode gluons and one KK gluon vanishes~\cite{Lillie:2007yh}). The mass scale of the first KK modes are, to zeroth approximation, $\pi / L$, and their coupling to the SM fermion $\psi_i$ (after diagonalizing the kinetic terms) is given by
\beq
g_{\psi_i \psi_i V_{1}} = \frac{C_{\psi_i}^{2}}{L} \left( \int^L_0 dy \  e^{-2M_{\Psi}y} f_{V,1}(y) + \alpha^{\psi_i} + \alpha_{0,ii}^{\psi_i} e^{-2M_{\Psi}L} f_{V,1}(L) \right).
\label{eq:gppg}
\eeq

\begin{figure}
\centering 
\includegraphics[width=.45\textwidth]{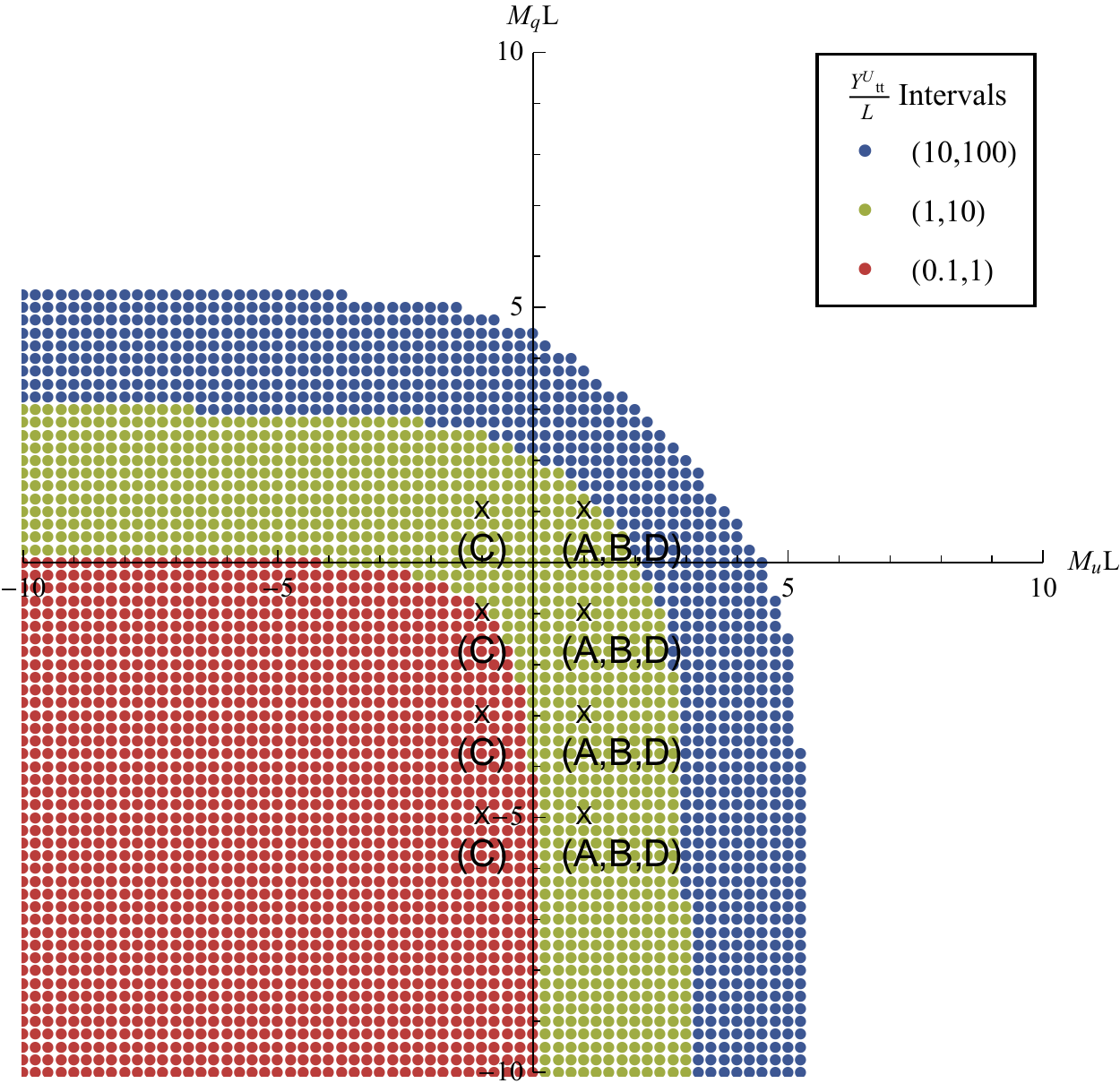}
\hfill
\includegraphics[width=.45\textwidth]{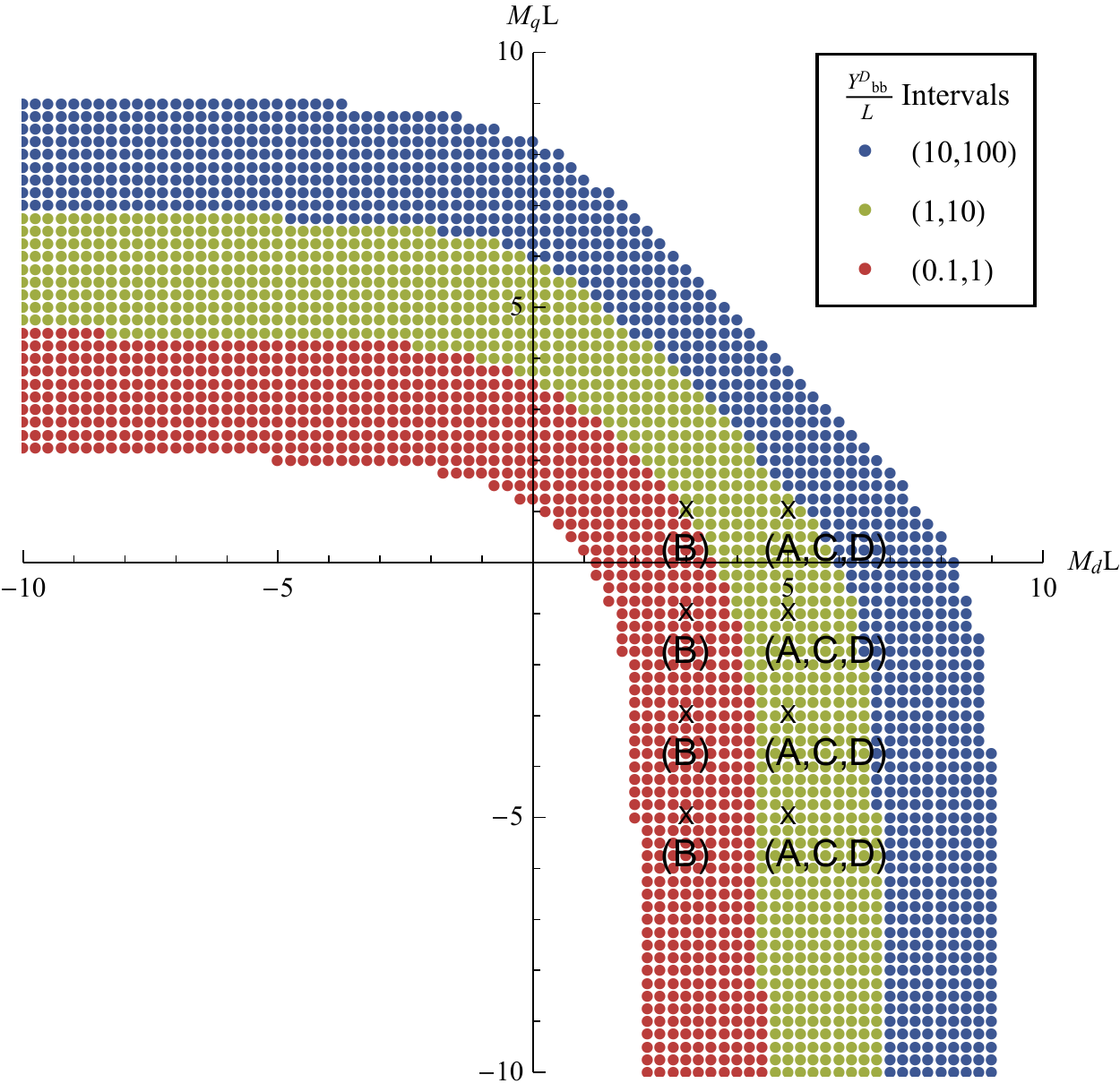}
	\caption{The median values of the $Y^U_{0,tt}/L$ and $Y^D_{0,bb}/L$ distributions chosen to reproduce the correct top and bottom mass respectively, for a range of choices of $M_u L$, $M_d L$ and $M_q L$, after 100,000 random trials at each point with $\mathcal{O}(1)$ symmetric BLKT coefficients as inputs. The sets of benchmark points A through D for several values of $M_{q} L$ are also shown.}
	\label{fig:YQ}
\end{figure}

In the same vein as the procedure for the leptons (equation ~\ref{eq:yukcpl}), we can determine the Yukawa couplings for the quarks after diagonalizing and normalizing their kinetic terms:
\begin{equation}
\begin{aligned}
Y^U_{ij} &\equiv \frac{Y^U_{0,ij}}{L} C_q C_{u} e^{-(M_q + M_{u})L}, \\
Y^D_{ij} &\equiv \frac{Y^D_{0,ij}}{L} C_q C_{d} e^{-(M_q + M_{d})L},
\end{aligned}
\label{eq:udyuks}
\end{equation}
where for given values of bulk masses for the quark fields, the diagonal entries $Y^U_{0,ii}/L$ and $Y^D_{0,ii}/L$ are chosen such that the correct SM quark masses are obtained. Similar to figure~\ref{fig:Y5L}, we show in figure~\ref{fig:YQ} the median values of the $Y^U_{0,tt}/L$ and $Y^D_{0,bb}/L$ distributions as a function of the bulk quark mass parameters $M_u L$, $M_d L$, and $M_q L$. Since the top mass is large, having $Y^U_{0,ii}/L$ be order one requires a negative bulk mass for one (or both) of $u$ or $q$. Thus, one (or both) of these quark profiles peaks at the Higgs brane, in which case the Higgs brane BLKTs are not suppressed. Unlike the leptons however, the quarks have no interactions on the FDM brane, and due to gauge invariance, the BLKT's are diagonal in the same basis as the kinetic terms. Therefore there is no basis mismatch giving rise to quark flavor changing processes, in addition to those already present in the SM. 

\begin{table}
\begin{center}
\begin{tabular}{c|c|c|c|c}
Benchmark point & $M_e L$ & $M_\ell L$ & $M_u L$ & $M_d L$\\
\hline
A & 8 & 1 & 1 & 5\\
B & 8 & 1 & 1 & 3\\
C & 8 & 1 & -1 & 5\\
D & 9 & -1 & 1 & 5
\end{tabular}
\end{center}
\caption{Four benchmarks for bulk masses to be used in our study of KK resonance bounds. For each of these choices, we will also vary $M_q L$.}
\label{tab:ABCD}
\end{table}

In order to demonstrate how the KK resonance bounds depend on the model parameters (in particular, the bulk masses), we choose four benchmarks, listed in table~\ref{tab:ABCD}. These points (for several values of $M_{q} L$) are also shown in figures~\ref{fig:Y5L} and \ref{fig:YQ}. Benchmark A is chosen such that in addition to the model being consistent with X-ray constraints, the nominal values of $Y^D_{0,bb}/L$ and $Y^U_{0,tt}/L$ are close to 1. Benchmarks B, C and D represent variations around benchmark A, where in benchmark B a lower value of $M_{d} L$ is considered (resulting in smaller values of $Y^D_{0,bb}/L$ at fixed $M_q$), in benchmark C a lower value of $M_{u} L$ is considered (resulting in smaller values of $Y^U_{0,tt}/L$ at fixed $M_q$), and where benchmark D is chosen to be in even less tension with X-ray constraints than benchmark A, while keeping the value of $Y^L_{0,\tau\tau}$ roughly the same.

\begin{figure}
\centering 
\includegraphics{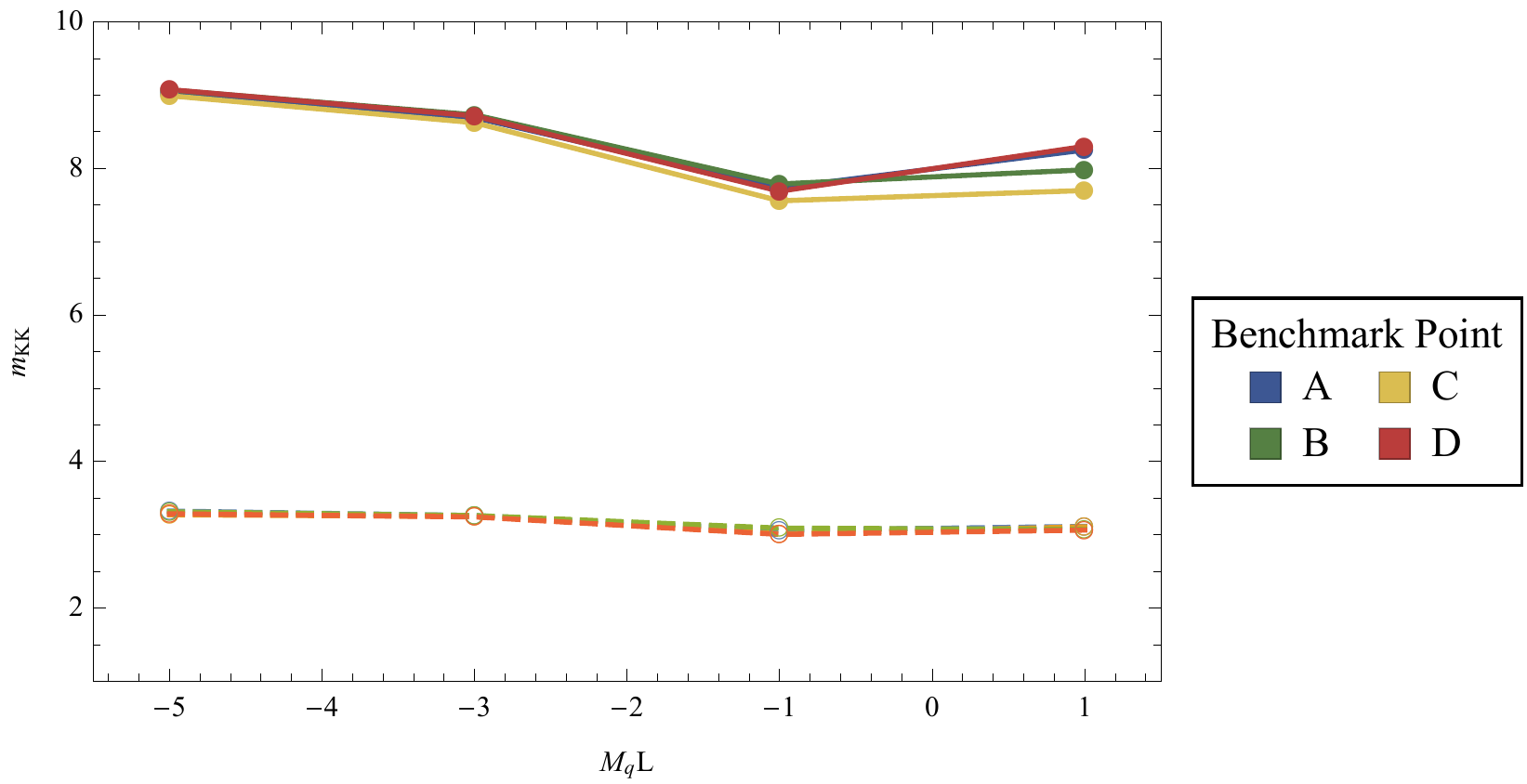}
	\caption{Constraints on the mass of the KK-$Z$ (dashed lines) and the KK gluon (solid lines) for the benchmarks A through D and for several values of $M_q L$, based on dilepton~\cite{Aad:2019fac} and dijet resonance searches~\cite{CMS:2019oju}, respectively.}
	\label{fig:KKGZ}
\end{figure}

In figure~\ref{fig:KKGZ}, we plot the constraints on the mass of the KK $Z$-boson and the KK gluon (first KK mode in each case), based respectively on dilepton resonance searches~\cite{Aad:2019fac} and dijet resonance searches~\cite{CMS:2019oju}, as we vary $M_{q} L$ for each of the four benchmarks A through D. The limits are extracted by performing a Monte Carlo scan over the BLKT coefficients (quarks as well as leptons) as before, and the exclusion curve corresponds to 95\% of the trials resulting in resonant production cross sections consistent with the experimental bounds (1000 trials per point). There are two main takeaway points from this figure, namely that the KK gluon bound is the dominant one, pushing the KK scale $\pi / L$ to values close to 10~TeV, and that the resonance bounds are fairly insensitive to the bulk mass parameters. Therefore in this paper we have used 10~TeV as a lower bound on the KK scale, but we remind the reader that the KK scale could in principle be much larger. Finally, with the KK scale at or above 10~TeV, no KK modes can be pair produced at the LHC, leading to no additional constraints.

\section{Conclusions}
\label{sec:conclusions}

We have studied a UV completion of lepton FDM in a flat extra dimension, where the DM and mediator fields, and the Higgs field live on branes on opposite ends of the extra dimension. With this setup, lepton flavor violating processes only arise as a result of the misalignment of bases which diagonalize the interactions on the Higgs and the FDM branes, and their size can be controlled by the lepton profiles along the extra dimension, which is achieved by an appropriate choice of bulk masses. 

Relic abundance, direct and indirect detection constraints can be satisfied as long as $m_{\chi}\gsim 300~$GeV, and $\lambda\sim \mathcal O(1)$. Due to the global flavor symmetry setup, the DM flavors are very nearly degenerate in mass, with $\chi_{e}$ lighter than $\chi_{\mu}$ by $\mathcal O$(10)eV and lighter than $\chi_{\tau}$ by $\mathcal O$(1)keV. The heavier flavors decay to $\chi_{e}$ via dipole transitions over very long lifetimes. The $\chi_{\tau}$ lifetime is constrained to be longer than the age of the universe, which puts an upper limit on the off-diagonal entries of the coupling matrix $\delta\lambda\lsim \mathcal O(10^{-6})$. In terms of model parameters, this constraint translates to $M_{e} L\gsim 8$. In this parameter range, lepton flavor violating processes such as $\mu\to e\gamma$ remain significantly below experimental constraints. Collider searches push the KK scale $\pi / L$ to 10~TeV or above, and give no additional constraints from the pair production of the mediator $\phi$ beyond the region that is excluded by direct detection.

Our study shows that in lepton-FDM models, a flavor structure that is consistent with constraints from flavor-changing processes can arise from an extra dimensional setup, with a broad range of parameters where all experimental constraints are satisfied. The mechanism in our setup that ameliorates flavor related constraints is fairly general, and a similar construction could help address flavor related issues in other extensions of the Standard Model as well. While we have taken the extra dimension in our study to be flat, it would also be interesting to study whether there may be qualitative changes in the phenomenology in a warped extra dimensional setup.

\section*{Acknowledgements}

We are grateful to Zackaria Chacko for invaluable ideas, insights and discussions. We also thank Prateek Agrawal, Cynthia Trendafilova and Christopher Verhaaren for helpful comments and discussions.
The research of the authors is supported by the National Science 
Foundation Grant Numbers PHY-1620610 and PHY-1914679.


\bibliographystyle{JHEP}
\bibliography{bib_FDM_xdim}{}
\end{document}